\documentclass[%
prl,
 reprint,
 superscriptaddress,
 amsmath,amssymb,
 aps,showkeys,floatfix
]{revtex4-2}

\usepackage{graphicx}
\usepackage{dcolumn}
\usepackage{bm}
\usepackage{xcolor}
\usepackage[mathlines]{lineno}

 \usepackage{tikz}
\usetikzlibrary{decorations}
\usetikzlibrary{decorations.pathmorphing}
\usetikzlibrary{shapes,backgrounds}
\usetikzlibrary{decorations.pathreplacing}
\usetikzlibrary{decorations.markings}
\usetikzlibrary{arrows,matrix,snakes}
\usepackage{mathtools}
\usetikzlibrary{trees}
\usetikzlibrary{graphs}

\tikzset{->-/.style n args={2}{decoration={markings, mark=at position #1 with {\arrow{#2}}},postaction={decorate}}}

\begin{document}

\preprint{APS/123-QED}

\title{First Measurement of Deeply Virtual Compton Scattering on the Neutron with Detection of the Active Neutron}

\author{A. Hobart}
\affiliation{\ORSAY}
\author{S. Niccolai}
\affiliation{\ORSAY}
\author{M. Čuić} \affiliation{Department of Physics, Faculty of Science, University of Zagreb, 10000 Zagreb, Croatia}
\author{K. Kumerički}\affiliation{Department of Physics, Faculty of Science, University of Zagreb, 10000 Zagreb, Croatia}


\newcommand*{\ANL}{Argonne National Laboratory, Argonne, Illinois 60439}
\newcommand*{\ANLindex}{1}
\newcommand*{\CSUDH}{California State University, Dominguez Hills, Carson, California 90747}
\newcommand*{\CSUDHindex}{2}
\newcommand*{\CANISIUS}{Canisius College, Buffalo, New York 14208}
\newcommand*{\CANISIUSindex}{3}
\newcommand*{\SACLAY}{IRFU, CEA, Universit\'{e} Paris-Saclay, F-91191 Gif-sur-Yvette, France}
\newcommand*{\SACLAYindex}{4}
\newcommand*{\CNU}{Christopher Newport University, Newport News, Virginia 23606}
\newcommand*{\CNUindex}{5}
\newcommand*{\UCONN}{University of Connecticut, Storrs, Connecticut 06269}
\newcommand*{\UCONNindex}{6}
\newcommand*{\DUKE}{Duke University, Durham, North Carolina 27708-0305}
\newcommand*{\DUKEindex}{7}
\newcommand*{\DUQUESNE}{Duquesne University, 600 Forbes Avenue, Pittsburgh, Pennsylvania 15282 }
\newcommand*{\DUQUESNEindex}{8}
\newcommand*{\FU}{Fairfield University, Fairfield, Connecticut 06824}
\newcommand*{\FUindex}{9}
\newcommand*{\FERRARAU}{Universita' di Ferrara, 44121 Ferrara, Italy}
\newcommand*{\FERRARAUindex}{10}
\newcommand*{\FIU}{Florida International University, Miami, Florida 33199}
\newcommand*{\FIUindex}{11}
\newcommand*{\GWUI}{The George Washington University, Washington, DC 20052}
\newcommand*{\GWUIindex}{12}
\newcommand*{\GSIFFN}{GSI Helmholtzzentrum fur Schwerionenforschung GmbH, D-64291 Darmstadt, Germany}
\newcommand*{\GSIFFNindex}{13}
\newcommand*{\INFNFE}{INFN, Sezione di Ferrara, 44100 Ferrara, Italy}
\newcommand*{\INFNFEindex}{14}
\newcommand*{\INFNFR}{INFN, Laboratori Nazionali di Frascati, 00044 Frascati, Italy}
\newcommand*{\INFNFRindex}{15}
\newcommand*{\INFNGE}{INFN, Sezione di Genova, 16146 Genova, Italy}
\newcommand*{\INFNGEindex}{16}
\newcommand*{\INFNRO}{INFN, Sezione di Roma Tor Vergata, 00133 Rome, Italy}
\newcommand*{\INFNROindex}{17}
\newcommand*{\INFNTUR}{INFN, Sezione di Torino, 10125 Torino, Italy}
\newcommand*{\INFNTURindex}{18}
\newcommand*{\INFNPAV}{INFN, Sezione di Pavia, 27100 Pavia, Italy}
\newcommand*{\INFNCAT}{INFN - Sezione di Catania, 95123 Catania, Italy}
\newcommand*{\INFNPAVindex}{19}
\newcommand*{\ORSAY}{Universit\'{e} Paris-Saclay, CNRS/IN2P3, IJCLab, 91405 Orsay, France}
\newcommand*{\ORSAYindex}{20}
\newcommand*{\Juelich}{Institute fur Kernphysik (Juelich), Juelich, Germany}
\newcommand*{\Juelichindex}{21}
\newcommand*{\JMU}{James Madison University, Harrisonburg, Virginia 22807}
\newcommand*{\JMUindex}{22}
\newcommand*{\KNU}{Kyungpook National University, Daegu 41566, Republic of Korea}
\newcommand*{\KNUindex}{23}
\newcommand*{\LAMAR}{Lamar University, 4400 MLK Blvd, PO Box 10046, Beaumont, Texas 77710}
\newcommand*{\LAMARindex}{24}
\newcommand*{\MIT}{Massachusetts Institute of Technology, Cambridge, Massachusetts 02139-4307}
\newcommand*{\MITindex}{25}
\newcommand*{\MISS}{Mississippi State University, Mississippi State, Mississippi 39762-5167}
\newcommand*{\MISSindex}{26}
\newcommand*{\UNH}{University of New Hampshire, Durham, New Hampshire 03824-3568}
\newcommand*{\UNHindex}{27}
\newcommand*{\NMSU}{New Mexico State University, PO Box 30001, Las Cruces, New Mexico 88003, USA}
\newcommand*{\NMSUindex}{28}
\newcommand*{\NSU}{Norfolk State University, Norfolk, Virginia 23504}
\newcommand*{\NSUindex}{29}
\newcommand*{\OHIOU}{Ohio University, Athens, Ohio  45701}
\newcommand*{\OHIOUindex}{30}
\newcommand*{\ODU}{Old Dominion University, Norfolk, Virginia 23529}
\newcommand*{\ODUindex}{31}
\newcommand*{\JLUGiessen}{II Physikalisches Institut der Universitaet Giessen, 35392 Giessen, Germany}
\newcommand*{\JLUGiessenindex}{32}
\newcommand*{\RPI}{Rensselaer Polytechnic Institute, Troy, New York 12180-3590}
\newcommand*{\RPIindex}{33}
\newcommand*{\URICH}{University of Richmond, Richmond, Virginia 23173}
\newcommand*{\URICHindex}{34}
\newcommand*{\ROMAII}{Universita' di Roma Tor Vergata, 00133 Rome, Italy}
\newcommand*{\ROMAIIindex}{35}
\newcommand*{\MSU}{Skobeltsyn Institute of Nuclear Physics, Lomonosov Moscow State University, 119234 Moscow, Russia}
\newcommand*{\MSUindex}{36}
\newcommand*{\SCAROLINA}{University of South Carolina, Columbia, South Carolina 29208}
\newcommand*{\SCAROLINAindex}{37}
\newcommand*{\TEMPLE}{Temple University,  Philadelphia, Pennsylvania 19122}
\newcommand*{\TEMPLEindex}{38}
\newcommand*{\JLAB}{Thomas Jefferson National Accelerator Facility, Newport News, Virginia 23606}
\newcommand*{\JLABindex}{39}
\newcommand*{\UTFSM}{Universidad T\'{e}cnica Federico Santa Mar\'{i}a, Casilla 110-V Valpara\'{i}so, Chile}
\newcommand*{\UTFSMindex}{40}
\newcommand*{\INSUBRIA}{Universit\`{a} degli Studi dell'Insubria, 22100 Como, Italy}
\newcommand*{\INSUBRIAindex}{41}
\newcommand*{\BRESCIA}{Universit`{a} degli Studi di Brescia, 25123 Brescia, Italy}
\newcommand*{\BRESCIAindex}{42}
\newcommand*{\UCR}{University of California Riverside, 900 University Avenue, Riverside, California 92521, USA}
\newcommand*{\UCRindex}{43}
\newcommand*{\GLASGOW}{University of Glasgow, Glasgow G12 8QQ, United Kingdom}
\newcommand*{\GLASGOWindex}{44}
\newcommand*{\YORK}{University of York, York YO10 5DD, United Kingdom}
\newcommand*{\YORKindex}{45}
\newcommand*{\VIRGINIA}{University of Virginia, Charlottesville, Virginia 22901}
\newcommand*{\VIRGINIAindex}{46}
\newcommand*{\YEREVAN}{Yerevan Physics Institute, 375036 Yerevan, Armenia}
\newcommand*{\YEREVANindex}{47}
 
\newcommand*{\NOWINFNGE}{INFN, Sezione di Genova, 16146 Genova, Italy}
\newcommand*{\NOWJLAB}{Thomas Jefferson National Accelerator Facility, Newport News, Virginia 23606}

\author {P.~Achenbach} 
\affiliation{\JLAB}
\author {J.S.~Alvarado} 
\affiliation{\ORSAY}
\author {W.R.~Armstrong} 
\affiliation{\ANL}
\author {H.~Atac} 
\affiliation{\TEMPLE}
\author {H.~Avakian} 
\affiliation{\JLAB}
\author{L.~Baashen}
\altaffiliation[Current address: ]{King Saud University in Riyadh, Saudi Arabia}
\affiliation{\FIU}
% ...

\collaboration{The CLAS Collaboration}

\date{\today}

\begin{abstract}
Measuring Deeply Virtual Compton Scattering on the neutron is one of the necessary steps to understand the structure of the nucleon in terms of Generalized Parton Distributions (GPDs). Neutron targets play a complementary role to transversely polarized proton targets in the determination of the GPD $E$. This poorly known and poorly constrained GPD is essential to obtain the contribution of the quarks' angular momentum to the spin of the nucleon. DVCS on the neutron was measured for the first time selecting the exclusive final state by detecting the neutron, using the Jefferson Lab longitudinally polarized electron beam, with energies up to 10.6 GeV, and the CLAS12 detector. The extracted beam-spin asymmetries, combined with DVCS observables measured on the proton, allow a clean quark-flavor separation of the imaginary parts of the GPDs $H$ and $E$. 
\end{abstract}

\maketitle

Understanding the structure of 
the nucleon in terms of quarks and gluons, collectively called partons, is one of the main challenges of hadronic physics. The formalism of Generalized Parton Distributions (GPDs) \cite{muller,Ji1,Ji2,Radyushkin,collins,goeke,diehl,belitsky} provides a universal description of the partonic structure of the nucleon. GPDs 
correlate partons in different quantum states, and can be interpreted as the spatial distributions in the transverse plane of partons carrying a given longitudinal momentum fraction. The simultaneous knowledge of longitudinal momentum and transverse position gives access to the angular momentum of quarks and gluons \cite{Ji1,Ji2}. Therefore, the determination of GPDs can clarify the so-called “spin crisis”, which ensued from the measurements \cite{spin_crisis} showing that the spins of the quarks contribute to only 20-30\% of the nucleon's spin. 

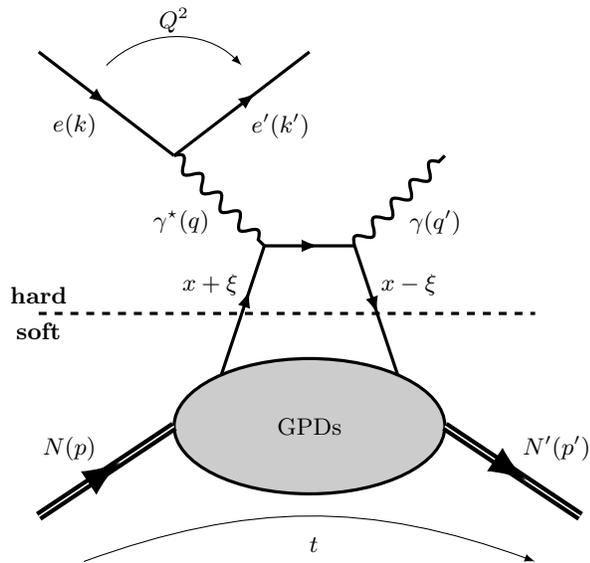
\begin{figure}[h!]
\begin{center}
\begin{tikzpicture}[very thick,>\relax=latex,scale=1.2]

\draw [->-={0.5}{>}]  (-1.5,1.15) -- (0,0) node [below left,pos=0.5]{$e(k)$}; 
\draw [->-={0.6}{>}] (0,0) -- (1.5,1.15) node [below right,pos=0.5]{$e^\prime (k^\prime)$}; 

\draw[->,thin] (-0.75,1) to [bend left=45] (0.75,1) ;
\node [above] at (0,1.25) {$Q^2$};

\draw[snake=coil,segment aspect=0] (0,0) -- (1,-1) node [below left,pos=0.5]{$\gamma^\star(q)$}; 

\draw [->-={0.6}{>}]  (1,-1) -- (2,-1) ;
\draw [->-={0.45}{<}]  (1,-1) -- (0.5,-2.5) ;
\node [above] at (0.4,-1.65) {$x+\xi$};
\draw[->-={0.5}{>}] (2,-1) -- (2.5,-2.5) ;
\node [above] at (2.6,-1.65) {$x-\xi$};

\draw [dashed]  (-1.5,-1.75) -- (4,-1.75) ;
\node[above] at (-1.5,-1.75)  {\bf{hard}};
\node[below] at (-1.5,-1.75)  {\bf{soft}};

\draw[snake=coil,segment aspect=0] (2,-1) -- (3,0) node [below  right, pos=0.5]{$\gamma(q^\prime)$}; 
\draw [fill=gray!40](1.5,-3) ellipse (1.5cm and 0.75cm);
\node[above] at (1.5,-3.2)  {GPDs};

\draw[->,thin] (-1,-4.5) to [bend left=20] (4,-4.5) ;
\node[above] at (1.55,-4.5)  {$t$};

\draw [->-={0.6}{>},ultra thick, double]  (-1.5,-4) -- (0,-3) node [above left,  pos=0.5]{$N(p)$};
\draw [->-={0.6}{>},ultra thick, double]   (3,-3)--(4.5,-4)   node [above right, pos=0.5]{$N^\prime(p^\prime)$};

\end{tikzpicture}
\end{center}
\caption{The ``handbag'' diagram for the DVCS process on the nucleon $eN\to e'N'\gamma$. The four-vectors of the incoming/outgoing electrons, photons, and nucleons are denoted by $k$/$k'$, $q$/$q'$, and $p/p'$, respectively. $t=(p-p')^2$ and $\xi$ is proportional to the Bjorken variable $x_B$ ($\xi\simeq\frac{x_B}{2-x_B}$, where $x_B=\frac{Q^2}{2M\nu}$, $M$ is the nucleon mass, and $\nu=E_e-E_{e'}$). }  \label{fig:handbagDVCS}
\end{figure}

GPDs derive from the theory of the strong interaction, quantum chromodynamics (QCD), as they are the Fourier transforms of non-local and non-diagonal QCD operators. 
They are most easily accessed in the measurement of the exclusive leptoproduction of a photon (DVCS, which stands for Deeply Virtual Compton Scattering) or a meson on the nucleon, at sufficiently large $Q^2$, which is the virtuality of the photon emitted by the initial lepton ($Q^2=-(k-k')^2$, where $k$ and $k'$ are the momenta of the initial and final state leptons, respectively). Figure~\ref{fig:handbagDVCS} illustrates the leading-order diagram for DVCS, where QCD factorization is applied, splitting the process into the hard quark-photon scattering part, calculable in perturbative quantum electrodynamics (QED), and the soft nucleon-structure part. Considering only helicity-conserving processes and the quark sector, the soft structure of the nucleon is parametrized by four GPDs for each quark flavor: $H, \tilde H, E, \tilde E$, which depend, in leading-order and leading-twist QCD, upon three variables: $x$, 
$\xi$, and $t$. $x$, the average parton momentum fraction, is not accessible experimentally in the DVCS process. $x+\xi$ and $x-\xi$ are the longitudinal
momentum fractions of the quarks, respectively, coming out from and going back into the nucleon. $t$ is the squared four-momentum transfer between the final and initial state nucleons.
Figure \ref{fig:handbagDVCS} and its caption illustrate the definitions of the relevant variables. 

DVCS shares the same final state with the Bethe-Heitler (BH) process, where a real photon is emitted by either the incoming or the scattered electron. At the cross-section level BH is typically larger than DVCS, but information on the latter can be obtained by extracting the DVCS/BH interference term, and exploiting the fact that the amplitude from BH can be computed. Spin-dependent asymmetries, which at leading-twist depend mainly on the interference term, can then be connected to linear combinations of real and imaginary parts of Compton Form Factors (CFFs), defined for a generic GPD $F$ as \cite{belitski}
\begin{equation}\label{cff_definition_0}
\Re{\rm e}{\cal F} = {\cal P}\int_{0}^{1}dx\left[\frac{1}{x-\xi}\pm\frac{1}{x+\xi}\right]\left[ F(x,\xi,t)\mp F(-x,\xi,t)\right]  
\end{equation}
\begin{equation}\label{cff_definition}
\Im{\rm m}\mathcal{F} = F(\xi,\xi,t)\mp F(-\xi,\xi,t),
\end{equation}
where $\cal{P}$ is the principal value of the integral, and the top and bottom signs 
apply, respectively, to the unpolarized GPDs ($H$, $E$) and to the polarized GPDs ($\tilde{H}$, $\tilde{E}$).
Measuring GPDs is a complex 
task, calling for a long-term experimental program comprising the measurement of different observables~\cite{guidal1, belitsky}. 
Such a dedicated experimental program, mainly focused on a proton target, has been carried out worldwide, in particular at Jefferson Lab (JLab), with CLAS/CLAS12 and Hall A \cite{stepan,munoz,fx,shifeng,erin,pisano,hs,defurne_halla,defurne_nature,defurne}, and at HERA with HERMES~\cite{HERMES_2001,HERMES_2007,HERMES_2008,HERMES_2009,HERMES_2010,HERMES_2011_trans,HERMES_2012}, H1~\cite{H1_1, H1_2, H1_3}, and Zeus~\cite{zeus1,zeus2}, bringing strong constraints to the GPD $H$ and 
indications on the size and kinematic dependence of $\tilde{H}$. 

Measuring DVCS on both protons and neutrons is essential to carry out the quark-flavor separation of GPDs. Moreover, the beam-spin asymmetry (BSA, hereafter also denoted by $A_{LU}$ where $L$ indicates the longitudinally polarized beam and $U$ the unpolarized target) for DVCS on the neutron is strongly sensitive to the GPD $E$, which is poorly known and constrained. $E$ is of particular interest as it enters, along with $H$, in Ji's sum rule~\cite{Ji1,Ji2}
\begin{equation}\label{ji_sum_rule}
\sum_{q}\int_{-1}^{+1}dx \, x[H^{q}(x,\xi,t=0)+E^{q}(x,\xi,t=0)]=2\, J_{q},
\end{equation}
which links the total angular momentum $J_q$ carried by each quark $q$ to the sum of the second moments over $x$ of the GPDs $H$ and $E$.
In a first approximation, the BSA relates to the CFFs as \cite{guidal}
\begin{equation}
\label{eq:1}
A_{LU} \propto \sin\phi  \Im{\rm m}[F_1\mathcal{H}+\xi(F_1+F_2){\tilde{\mathcal{H}}}+kF_2\mathcal{E}],
\end{equation}
where $\phi$ is the angle between the lepton scattering and photon production planes, $F_1$ and $F_2$ are the Dirac and Pauli form factors, and $k=-t/4M^2$ with $M$ the nucleon's mass. Due to the different values of $F_1$ and $F_2$ for the proton and neutron, and to the small size of $\xi$, the BSA will be mainly sensitive to $\Im{\rm m}\mathcal{H}$ of the proton, if the target is a proton, and to $\Im{\rm m}\mathcal{E}$ of the neutron, if the target is a neutron. 

The importance of neutron targets in the DVCS phenomenology was established by a pioneering Hall A experiment \cite{malek} that was then repeated with higher statistics \cite{benali}. Both experiments measured polarized-beam cross section differences for DVCS off a neutron from a deuterium target by detecting the scattered electron and the DVCS/BH photon ($ed\to e'\gamma (np,d) $) and then subtracting data taken, in the same detection topology, on a hydrogen target ($ep\to e'\gamma (p)$). 

This paper presents results for the BSA of neutron-DVCS (nDVCS) from a deuterium target, $ed \to e'n\gamma(p)$. This is the first nDVCS measurement with detection of the recoil neutron. 

The experiment ran at JLab in Hall B, using the large acceptance spectrometer CLAS12~\cite{clas12} and the longitudinally polarized electron beam produced by the Continuous Electron Beam Accelerator Facility. An average beam polarization of $\sim$85\% was measured throughout the experiment using a M{\o}ller polarimeter. The 5-cm-long target was filled with unpolarized liquid-deuterium. The experiment ran 
 between February 2019 and January 2020 during three periods, collecting 
 an integrated luminosity of roughly 285 fb$^{-1}$. A quarter of the data was taken at a beam energy of 10.6~GeV, another 
quarter at 10.2~GeV, and half at 10.4~GeV.
Events with at least one 
electron, one photon, and one neutron were selected for the DVCS analysis. The electrons emitted at polar angles $7^\circ \lesssim \theta_e \lesssim 36^\circ$ were identified 
combining signals from the high-threshold Cherenkov counter \cite{htcc} and the electromagnetic calorimeters (ECAL) \cite{ecal}, and their kinematics were measured by the drift chambers \cite{dc}. The photons were identified and reconstructed by two 
electromagnetic calorimeters: the ECAL for 
$5^\circ \lesssim\theta_{\gamma}\lesssim 35^\circ$ and the Forward Tagger \cite{ft} for $2.5^\circ \lesssim \theta_{\gamma}\lesssim 4.5^\circ$. The neutrons were identified and their kinematics reconstructed either by the ECAL 
or by the Central Neutron Detector (CND) \cite{cnd}, 
conceived specifically for this experiment, and the Central Time-of-flight (CTOF) \cite{ctof}.

In the case where multiple final-state particles of the same type were detected in an event, all possible combinations 
were examined. The chosen combination was the one minimizing a 
$\chi^2$-like quantity calculated using variables related to the exclusive final state. 

To determine the selection criteria for the exclusivity, a GEANT-4 Monte-Carlo simulation of CLAS12 was used \cite{gemc}. An event generator for incoherent electroproduction of photons on deuterium was adopted, which produces either $ed\to e'n\gamma (p)$ or $ed\to e'p\gamma (n)$ 
events, proportionally to their relative cross sections, coming from the nDVCS/pDVCS and BH reactions~\cite{ahmed}. The DVCS amplitude is calculated according to the BMK formalism \cite{belitski}. The Fermi-motion distribution is implemented via the Paris potential \cite{paris}.

Several cuts were applied in order to ensure proper particle identification and select the relevant kinematic region for the DVCS reaction. Fiducial cuts were applied to remove the edges of the detector
. The electron momentum was required to be above 1~GeV. Only neutrons with momenta above 0.35~GeV were kept, in order to remove spectator-neutron events. The minimum photon energy was required to be 2~GeV. The cone angles formed by the electron and the neutrals, the photon or the neutron ($\theta_{e\gamma}$, $\theta_{en}$), were required to be bigger than $5^{\circ}$ to remove radiative
photons produced by the electrons while passing through the target and detector
materials, as well as those erroneously reconstructed neutral clusters identified as photons or neutrons while being part of the electron shower in the calorimeter.
Imposing $Q^2>1$~ GeV$^{2}$ and $W>2$~GeV ensured the applicability of the leading-twist GPD formalism and minimized contributions from nucleon resonances. 

\begin{figure}[hbt]
\includegraphics[width=0.9\columnwidth]{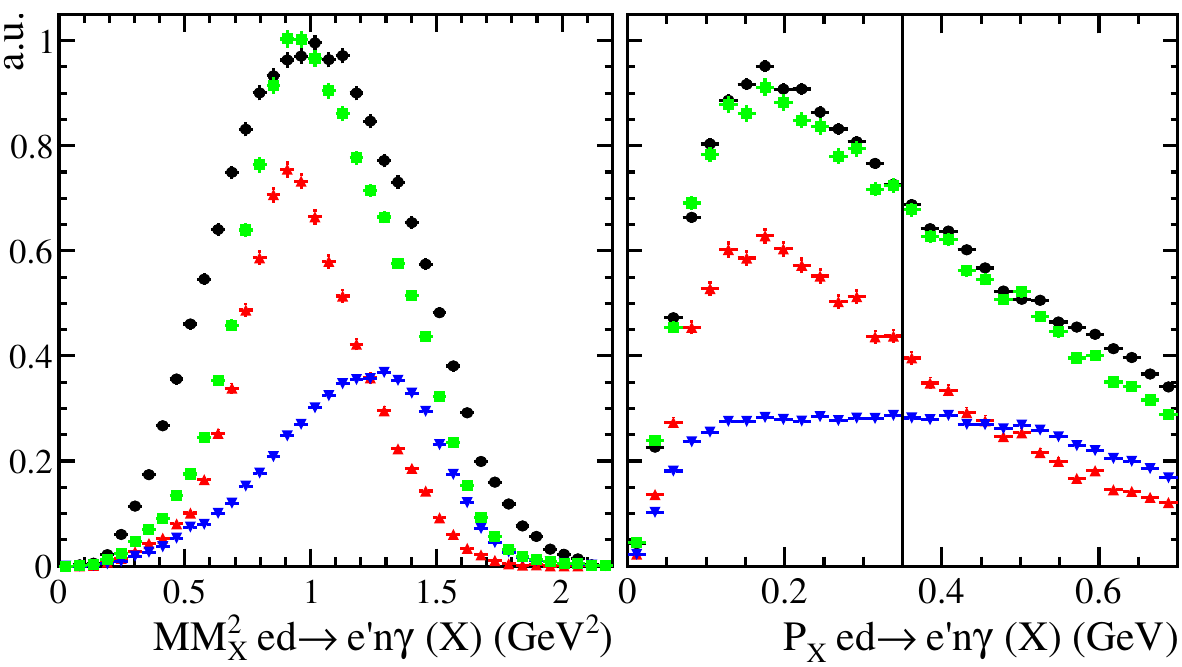}
\caption{Squared missing mass (left) and missing momentum (right) from $ed\to e'n \gamma X$. The line defines the applied cut on $P_X$. 
The data (black circles) are compared with simulations of neutron DVCS (red triangles) and of partially reconstructed $\pi^0$ background (blue upside-down triangles). The simulations are rescaled to match, approximately, the relative weights of each contribution to the data. The green squares are the sums of the two simulated contributions. }\label{fig:exclusivityndvcs}
\end{figure}

Exclusivity cuts were applied to select the $e'n\gamma(p)$ final state while minimizing the background coming from partially reconstructed $\pi^0$ decays from the $ed\to e'n\pi^0(p)$ reaction, where only one of the two photons from the $\pi^0$ decay was reconstructed and the event passed the DVCS selection cuts. Cuts on the missing masses of $X$ in the $en \to e'n\gamma X$ and $en\to e'nX$ reactions, and on the missing momentum of $X$ in $ed \to e'n\gamma X$ were imposed ($|MM^2_X(en \to e'n\gamma X)|<0.1$ GeV$^2$, $|MM^2_X(en \to e'n X)|<2.5$ GeV$^2$, $P_X(ed \to e'n\gamma X)<0.35$ GeV). A further cut was imposed on $\Delta\phi$ ($-1.5^\circ<\Delta\phi<0.75^\circ$), the difference between the two ways of computing the angle $\phi$ between the leptonic and hadronic planes (using the nucleon and the virtual photon and using the virtual and the real photon). A similar cut was applied on $\Delta t$ ($|\Delta t|<0.5$ GeV$^2$) the difference between the two ways to compute $t$, using either the scattered nucleon or the virtual and real photon. 
Finally, a cut on $\theta_{\gamma X}$, the cone angle between the detected $\gamma$ and the missing particle $X$ 
in $en\to e'n'X$
, was applied ($\theta_{\gamma X}<3^\circ$).
Figure~\ref{fig:exclusivityndvcs} shows the squared missing mass of $X$ in $ed \to e'n\gamma X$ and the missing momentum $P_X$ for the data and the simulations for DVCS and for $\pi^0$, after having applied the exclusivity cuts. 
The data still contain some background 
from partially reconstructed $\pi^0$ decays. 

\begin{figure}
  \centering
  \includegraphics[width=0.9\columnwidth]{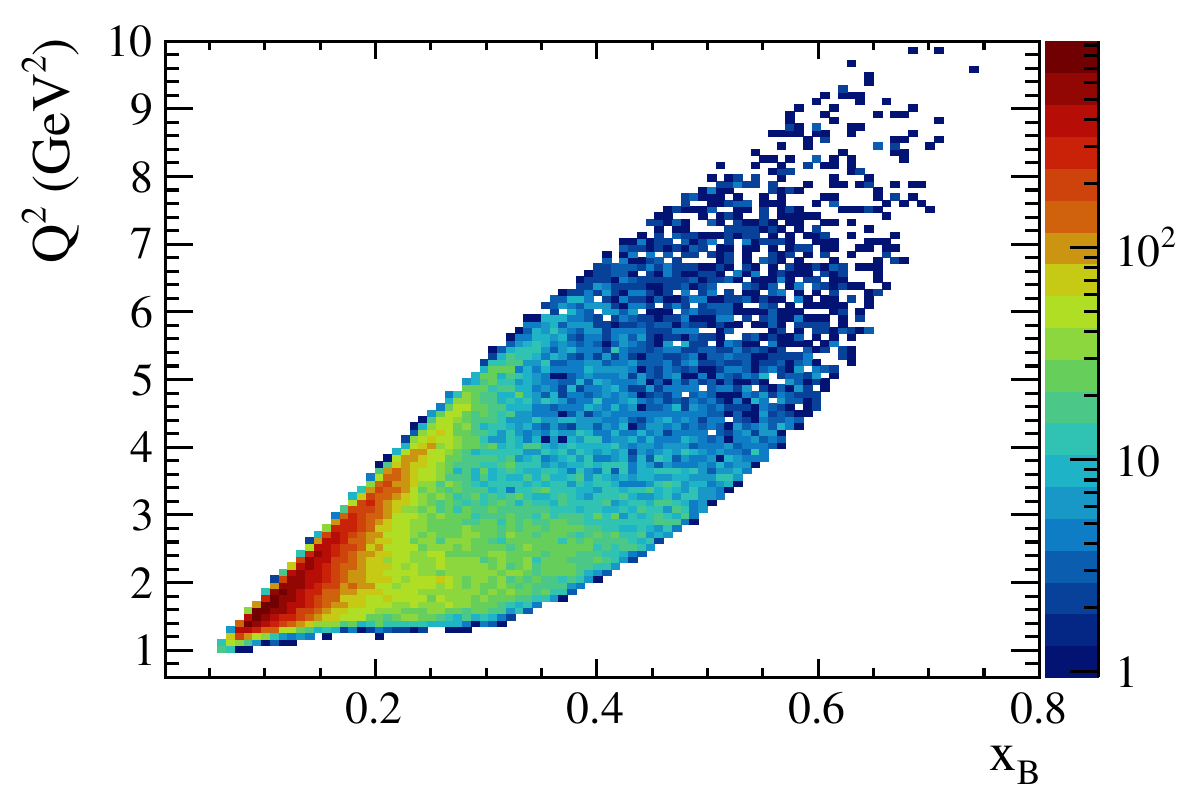}
  \caption{$Q^2$ versus $x_{B}$ for the nDVCS data sample with all selection cuts applied, showing the wide kinematic reach of CLAS12.}
  \label{fig:q2xbj}
\end{figure}

Due to 
inefficiencies in the Central Tracker \cite{cvt}, some protons were misidentified as neutrons. This background was reduced using a multivariate analysis technique (Boosted Decision Trees (BDT)~\cite{coadou}) that relied on low-level features from the CND and the CTOF, 
and on 
$\Delta\phi$. The remaining contamination from protons to the neutron sample was estimated to be $\sim 5\%$ and subtracted in the computation of the BSA.
Overall, 77580 events remained after all selections were applied. Figure~\ref{fig:q2xbj} shows the kinematic coverage in $Q^2$ and $x_{B}$ of the selected events. 

The $\pi^0$ contamination to the DVCS sample was evaluated and subtracted. 
First, the ratio, from simulations, of partially reconstructed $e'n\pi^0$(1$\gamma$) events passing the selection criteria for the DVCS process to fully reconstructed $e'n\pi^0$ events was computed. Multiplying this ratio by the number of reconstructed $e'n\pi^0$ events in the data yields the number of $e'n\pi^0$(1$\gamma$) events. This number was then subtracted from the yield of DVCS event candidates in each kinematic bin and helicity state. The $\pi^0$ contamination ranges from 10\% to 45\% depending on the kinematics. 

The BSA is obtained for each kinematic bin as
\begin{equation}\label{def_bsa}
A_{LU} = \dfrac{1}{P} \dfrac{N^+ - N^-}{N^+ + N^-},
\end{equation}
where $P$ is the average beam polarization and $N^{+(-)}$ is the yield of DVCS events for positive (negative) beam helicity after $\pi^0$ subtraction.
Radiative corrections were estimated according to Ref.~\cite{akushevich} and found to be negligible.

Various sources of systematic uncertainty on the BSA were studied. To obtain the systematic uncertainty due to the cut on the BDT classifier to remove the proton contamination and on the exclusivity cuts, variations around each chosen cut were made, and the differences between the resulting BSAs were taken as the systematic uncertainty. The systematic uncertainty on the beam polarization was the standard deviation of the polarization measured by the M{\o}ller polarimeter. The systematic uncertainty stemming from the merging of datasets with different beam energy was evaluated with 
a GPD-based model computing the DVCS-BH BSA. The systematic uncertainty induced by the $\pi^0$ subtraction method was estimated using a different method, relying on the statistical unfolding ~\cite{splot} of signal and background contributions to the $MM^2_{en\gamma X}$ spectrum (Fig.~\ref{fig:exclusivityndvcs}, left),  and comparing the obtained BSAs in each kinematic bin. The total systematic uncertainty was computed by summing all contributions in quadrature. It is, on average, $\sim 0.01$, and 
largely dominated by the uncertainty on the exclusivity cuts. 

\begin{figure}[htb]
  \begin{center}
  \includegraphics[width=0.98\columnwidth]{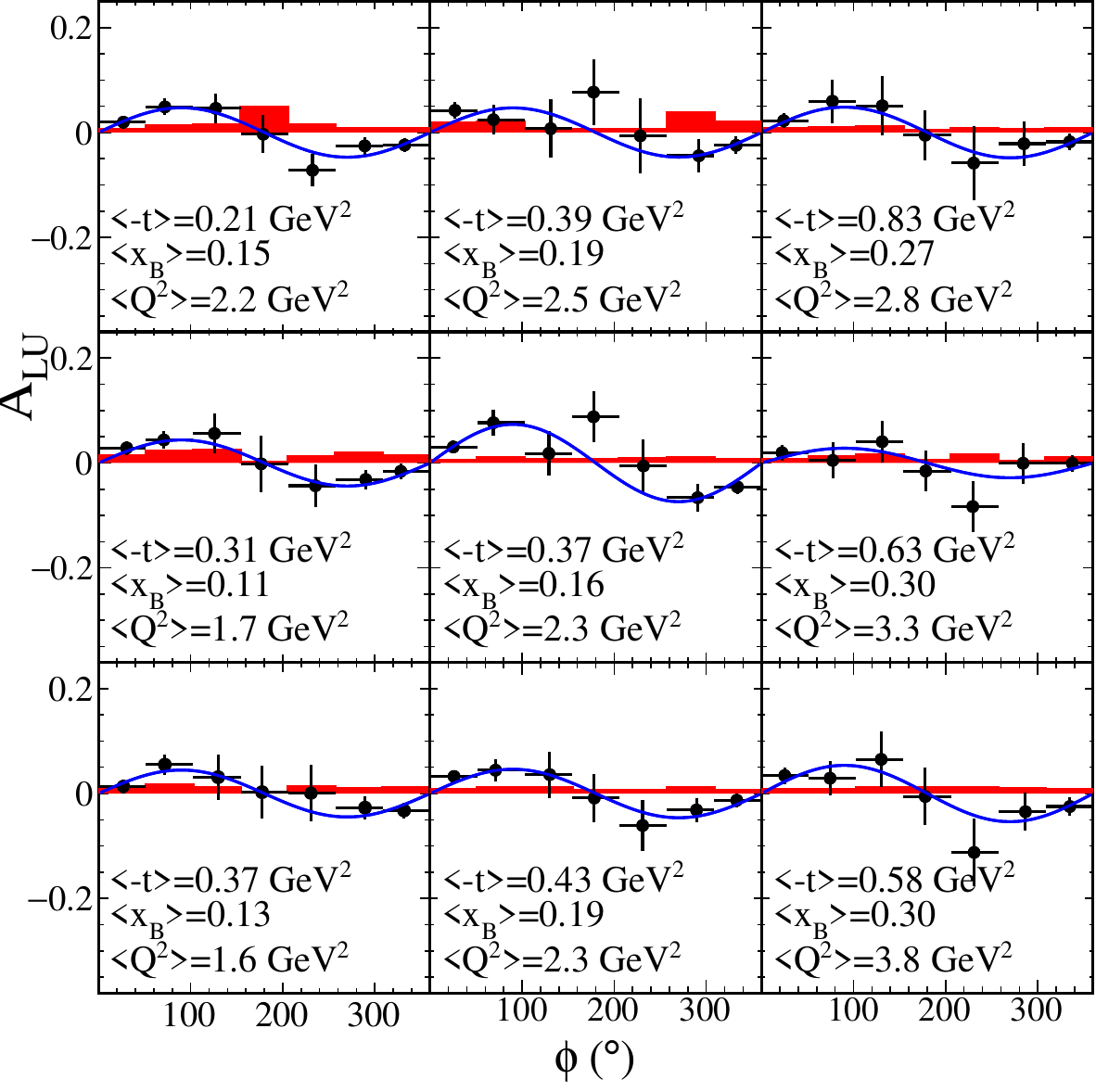}
  \caption {Beam-spin asymmetry for nDVCS versus $\phi$ for (top) three bins in $-t$ ([0, 0.3], [0.3, 0.5], and [0.5, 1.1]~GeV$^2$), (middle) three bins in $x_{B}$ ([0.05, 0.14], [0.14, 0.2], and [0.2, 0.6]), and (bottom) three bins in $Q^2$ ([1, 1.9], [1.9, 2.9], and [2.9, 6]~GeV$^2$). The error bars are statistical. The data are fit with the function $A_{LU}(90^{\circ})\sin\phi$. 
  The histogram shows the total systematic uncertainty.}
\label{fig_bsa}
  \end{center}
  \end{figure}

The BSA, which was extracted in bins of either $Q^2$, $x_{B}$, or $t$, is plotted as a function of $\phi$ in Fig.~\ref{fig_bsa}. 
It has the expected sinusoidal shape arising from the DVCS-BH interference, and is fitted by the function $A_{LU}(90^{\circ})\sin\phi$. Its amplitude is on the order of a few percent, about a factor of 4 smaller than the pDVCS amplitude measured at these same kinematics \cite{defurne}. The systematic uncertainty 
is consistently smaller than the statistical one. 

\begin{figure}[hbt]
\includegraphics[width=0.98\columnwidth]{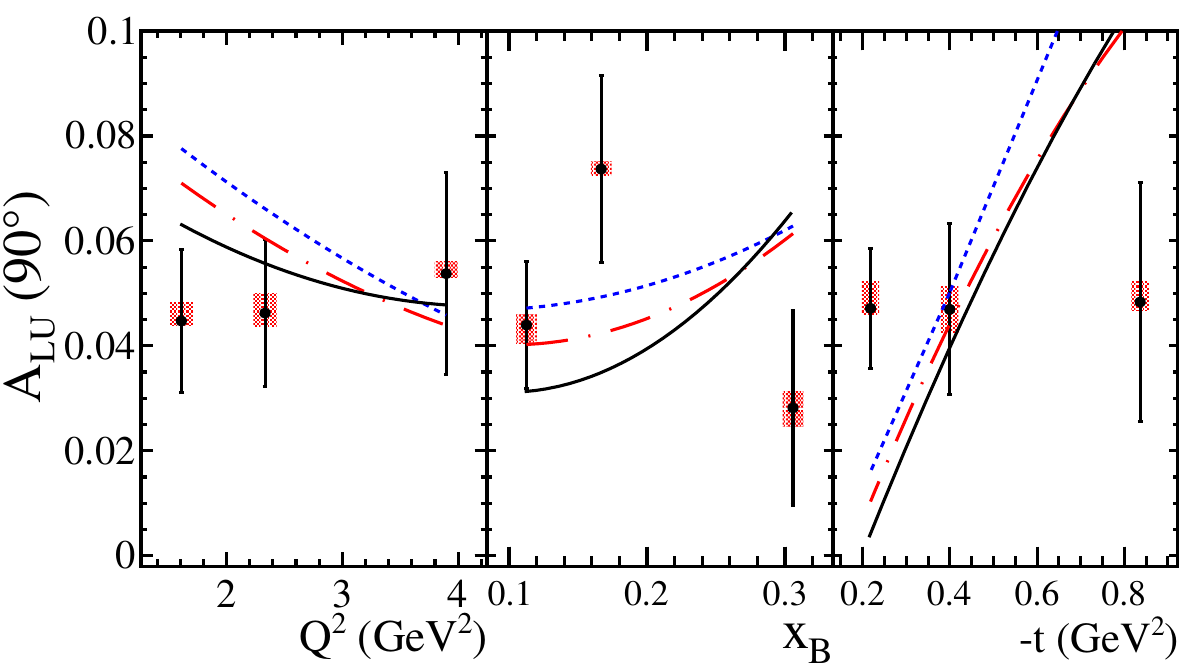}
\caption{\label{fig:sinphiq2} The $\sin\phi$ amplitude of $A_{LU}$ for nDVCS as a function of $Q^2$ (left), $x_B$ (middle), and $-t$ (right). The (red online) bands represent the systematic uncertainties. The VGG model \cite{vgg} predictions for three of the combinations of $J_u$ and $J_d$ yielding the best $\chi^2$ are compared to the data: solid (black online) line for $J_u=0.35$ $J_d=0.05$, dashed-dotted (red online) line for $J_u=-0.2$ $J_d=0.15$, and (blue online) dotted line for $J_u=-0.45$ $J_d=0.2$.
}\label{fig_alu}
\end{figure}

Figure~\ref{fig_alu} shows the amplitude $A_{LU}(90^{\circ})$ of the $\sin\phi$ fits to the BSA as a function of $Q^2$ (left), $x_B$ (middle), and $-t$ (right). The data are compared to predictions for DVCS on a free neutron of the VGG model \cite{vgg} for different values of the quark total angular momenta $J_u$ and $J_d$. The VGG model uses double distributions \cite{muller, Radyushkin} to parametrize the $(x, \xi)$ dependence of the GPDs, and Regge phenomenology for their $t$ dependence. The model curves are obtained at the average kinematics for $Q^2$, $x_B$, and $-t$, and setting $\phi$ at $90^{\circ}$. The values of $J_u$ and $J_d$ were varied in a grid of step 0.025 and range $\pm 1$, and the $\chi^2$ of each obtained model curve with the data points was computed. Three of the curves yielding the best $\chi^2$ are retained for Fig.~\ref{fig_alu}. Considering $\chi^2$ values within $3\sigma$ from the minimum, in the VGG framework the data favor $d$ quark angular momenta $0<J_d<0.2$, while no constraints can be imposed on $J_u$. 
The model does not reproduce the kinematic dependence of $A_{LU}(90^{\circ})$, predicting steeper variations, in particular for $-t$, than those displayed by the data. 

\begin{figure}[hbt]
\includegraphics[width=0.98\columnwidth]
{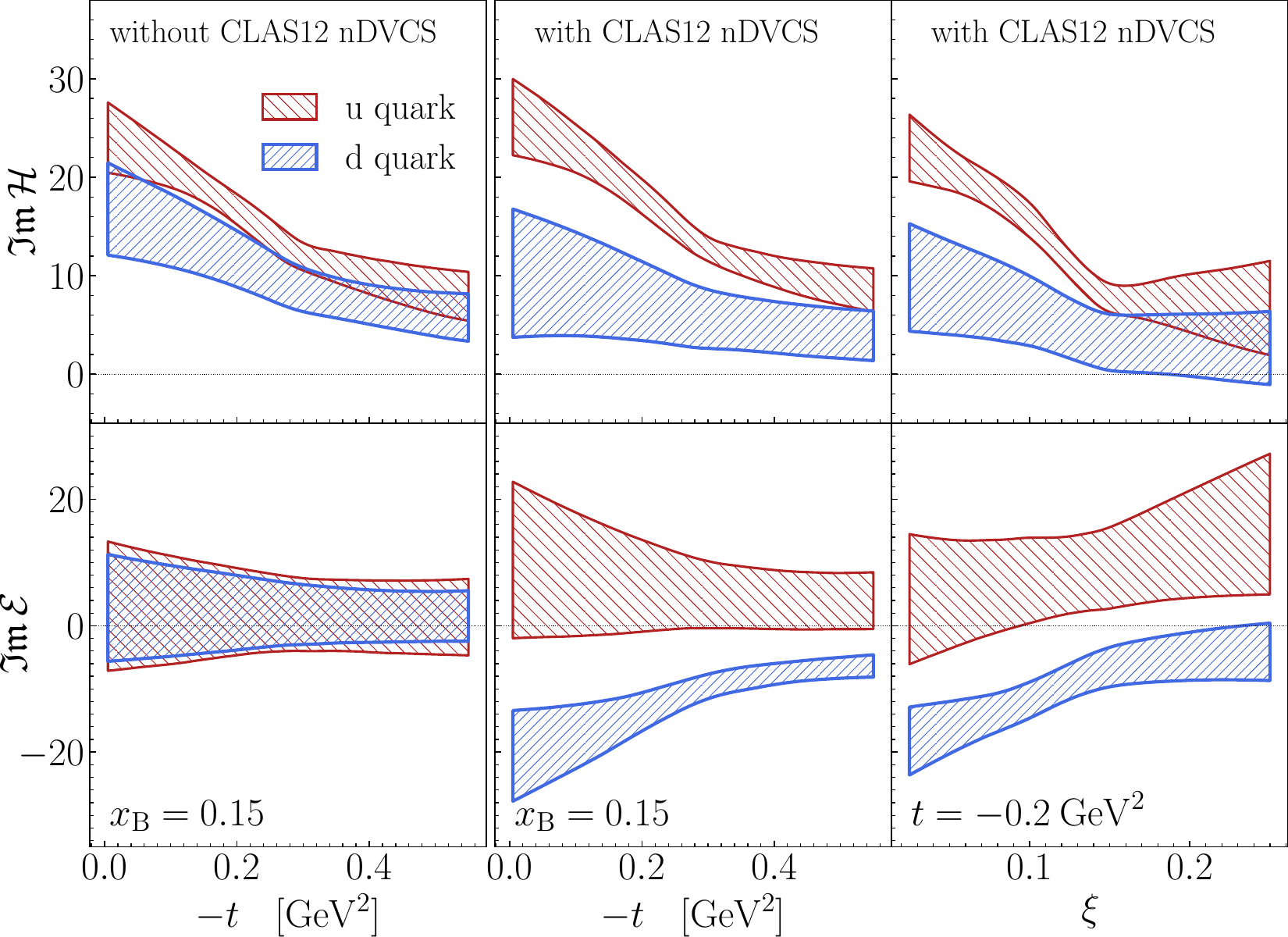}
\caption{\label{fig:flavsep} Extraction of up ($u$, coarser shading, red online) and down ($d$, finer shading, blue online) quark contributions to $\Im{\rm m}\mathcal{H}$ (top) and $\Im{\rm m}\mathcal{E}$ (bottom) as a function of $-t$ (left and middle) and $\xi$ (right). The leftmost column shows the extraction of the two CFFs without the CLAS12 nDVCS data, which are instead included in the other two columns.}
\end{figure}

The sensitivity of the CLAS12 nDVCS BSA to CFFs, in particular to $\Im{\rm m}\mathcal{E}$, was tested by including it 
in a non-biased fit method 
to extract CFFs \cite{kreso1}. In this method, the CFFs are parametrized as neural networks, with values at input representing the kinematical variables $x_B$, $Q^2$, and $t$, and values at output representing the imaginary or real parts of the CFFs. 
Figure~\ref{fig:flavsep} shows the up and down quark $\Im{\rm m}\mathcal{H}$ and $\Im{\rm m}\mathcal{E}$ CFFs, extracted by fits to old CLAS \cite{pisano, hs} and HERMES \cite{HERMES_2010,HERMES_2012} proton data, to recent CLAS12 proton data \cite{defurne} and to the neutron data reported here. The flavored models used for Fig.~\ref{fig:flavsep} have 200 trained neural nets to optimize the statistics. Note that the Hall A nDVCS data were not included in this study, in order to assess the impact of the present data alone. While the inclusion of the CLAS12 nDVCS data allows the flavor separation of $\Im{\rm m}\mathcal{H}$, the main new result is 
the flavor separation of $\Im{\rm m}\mathcal{E}$. 
The same CFF extraction method was previously used to attempt flavor separation of the CFFs of $H$ and $E$ \cite{kreso2} by combining pDVCS data and the Hall A nDVCS results \cite{benali}. While promising results were obtained for $\Re{\rm e}\mathcal{H}$ and $\Im{\rm m}\mathcal{H}$, the separation was not possible for the CFFs of $E$. The small systematic uncertainties of the CLAS12 nDVCS data, obtained mainly thanks to the detection of the neutron, and their wide kinematic coverage provide the necessary sensitivity for the flavor separation of $\Im{\rm m}\mathcal{E}$. 
%

The CLAS12 nDVCS data represent an important step towards the understanding of the contribution of the angular momentum of the quarks to the spin of the nucleon via Ji's sum rule, of which the GPD $E$ is an essential, yet poorly known, ingredient. The future increase in statistics of the nDVCS dataset, which will be achieved both by upgrades of the CLAS12 reconstruction software and by additional data, will allow better precision and 4-dimensional binning for the BSA, and thus a more accurate mapping of $\Im{\rm m}\mathcal{E}$. An ongoing analysis of this same dataset aims to extract cross sections for nDVCS, which are 
sensitive to $\Re{\rm e}\mathcal{E}$. Furthermore, CLAS12 
recently completed an experiment with a longitudinally polarized deuterium target, which will yield target-spin asymmetries and double-spin asymmetries for nDVCS. These observables, combined with the unpolarized-target ones, will contribute to constrain more CFFs of the neutron, hence to progress in the flavor separation of all 4 GPDs, and, consequently, to deepen our understanding of the properties of the nucleons in terms of their elementary constituents. 

We thank very much the staff of the Accelerator and the Physics Divisions at Jefferson Lab for making the experiment possible. This work
was supported in part by the U.S. Department of Energy (DOE) and National Science Foundation (NSF),
the Italian Istituto Nazionale di Fisica Nucleare (INFN), the French Centre National de la Recherche Scientifique (CNRS), the French Commissariat \`{a} l’Energie Atomique (CEA), the Scottish Universities Physics Alliance (SUPA), the National Research Foundation of Korea (NRF), the UK Science and Technology Facilities Council (STFC), the Chilean
Agencia Nacional de Investigacion y Desarrollo (ANID), and the Croatian Science
Foundation. This work was supported as well by the EU Horizon 2020 Research and Innovation Program under the STRONG-2020 Grant Agreement No. 824093 and the Marie Sklodowska-Curie Grant Agreement No. 101003460. The Southeastern Universities Research Association (SURA) operates the Thomas Jefferson National Accelerator Facility for the U.S. Department of Energy under Contract No. DE-AC05-06OR23177.

\end{document}